\documentstyle[12pt,epsfig,wrapfig]{article}
\renewcommand{\section}[1]{\vspace{6pt} \noindent\mbox{#1} \newline \noindent}
\renewcommand{\subsection}[1]{\vspace{6pt} \noindent\mbox{\underline{#1}} 
\newline \noindent}
\renewcommand{\subsubsection}[1]{\vspace{6pt} \noindent\mbox{\underline{#1}}
\noindent}

\newfont{\sansb}{cmssbx10}
\newfont{\sans}{cmss10}

\begin{document}
{\small OG 4.1.12 \vspace{-24pt}\\}

{\center \LARGE OBSERVATIONS OF PULSARS, PSR $1509-58$ AND PSR $1259-63$,
BY CANGAROO 3.8~m TELESCOPE
\vspace{6pt}\\}

T. Sako$^1$, S. A. Dazeley$^2$, P. G. Edwards$^3$, T. Hara$^4$, Y. Hayami$^5$,
S. Kamei$^5$, T. Kifune$^6$, R. Kita$^7$, T. Konishi$^8$, A. Masaike$^9$,
Y. Matsubara$^1$, T. Matsuoka$^1$, Y. Mizumoto$^{10}$, M. Mori$^{11}$,
H. Muraishi$^7$, Y. Muraki$^1$, T. Naito$^{12}$, K. Nishijima$^{13}$,
S. Ogio$^5$, J. R. Patterson$^2$, M. D. Roberts$^6$, G. P. Rowell$^2$,
K. Sakurazawa$^5$, R. Susukita$^{14}$, A. Suzuki$^8$, R. Suzuki$^5$,
T. Tamura$^{15}$, T Tanimori$^5$, G. J. Thornton$^2$, S. Yanagita$^7$,
T. Yoshida$^7$, and T. Yoshikoshi$^6$
\vspace{6pt}\\

{$^1Solar$-$Terrestrial$ $Environment$ $Laboratory$, $Nagoya$ $University$, 
$Nagoya$ $464$-$01$, $Japan$ \\
$^2Department$ $of$ $Physics$ $and$ $Mathematical$ $Physics$, $University$ 
$of$ $Adelaide$, \\
$South$ $Australia$ $5005$, $Australia$ \\
$^3Institute$ $of$ $Space$ $and$ $Astronomical$ $Science$, $Sagamihara$ 
$229$, $Japan$ \\
$^4Faculty$ $of$ $Commercial$ $Science$, $Yamanashi$ $Gakuin$ $University$, 
$Kofu$ $400$, $Japan$ \\
$^5Department$ $of$ $Physics$, $Tokyo$ $Institute$ $of$ $Technology$, 
$Tokyo$ $152$, $Japan$ \\
$^6Institute$ $for$ $Cosmic$ $Ray$ $Research$, $University$ $of$ $Tokyo$, 
$Tokyo$ $188$, $Japan$ \\
$^7Faculty$ $of$ $Science$, $Ibaraki$ $University$, $Mito$ $310$, $Japan$ \\
$^8Department$ $of$ $Physics$, $Kobe$ $Unibersity$, $Kobe$ $637$, $Japan$ \\
$^9Department$ $of$ $Physics$, $Kyoto$ $Unibersity$, $Kyoto$ $606$, $Japan$ \\
$^{10}National$ $Astronomical$ $Observatory$ $of$ $Japan$, $Tokyo$ $181$, 
$Japan$ \\
$^{11}Faculty$ $of$ $Education$, $Miyagi$ $University$ $of$ $Education$, 
$Sendai$ $980$, $Japan$ \\
$^{12}Department$ $of$ $Physics$, $Faculty$ $of$ $Science$, $University$ $of$ 
$Tokyo$, $Tokyo$ $113$, $Japan$ \\
$^{13}Department$ $of$ $Physics$, $Tokai$ $University$, $Hiratsuka$ $259$, 
$Japan$ \\
$^{14}Institute$ $of$ $Physical$ $and$ $Chemical$ $Research$, $Wako$ 
$351$-$01$, $Japan$ \\
$^{15}Faculty$ $of$ $Engineering$, $Kanagawa$ $University$, $Yokohama$ $221$, 
$Japan$ \vspace{-12pt}\\}

{\center ABSTRACT\\}
The data for VHE ($\sim$TeV) gamma rays from young gamma-ray pulsar 
PSR1509-58 observed in 1996 with the CANGAROO 3.8m \v{C}erenkov imaging 
telescope are presented, 
as well as the additional data from March to June of 1997. 
The high spin-down luminosity of the pulsar and the plerionic feature around the 
pulsar observed with radio and X-rays suggest that VHE gamma-ray emission 
is quite likely above the sensitivity of the CANGAROO telescope. 
The CANGAROO results on other pulsars, such as PSR1259-63, 
are also presented. PSR1259-63 is a highly eccentric X-ray binary system, 
which includes a high mass Be companion star,
and a preliminary analysis on the data taken 4 months after the periastron 
in 1994 suggests emission of VHE gamma rays.

\setlength{\parindent}{1cm}
\section{INTRODUCTION}
Extensive efforts have been made for detecting VHE gamma rays
from galactic objects which are considered to
be VHE gamma ray emitters such as pulsars, supernova remnants and
 EGRET sources.
The Crab, PSR~B1706-44 and the Vela pulsars are found to be
VHE gamma ray sources by using imaging \v Cerenkov telescope,
and, as these examples indicate,
 a young rotation-powered pulsar is a likely accelerator
of non-thermal energetic particles. A shock mechanism may occur between
the pulsar wind and circumstellar matter to form a pulsar nebula, where
particle acceleration and/or randomization of the pulsar wind flow of
relativistic particles takes place.
Non-thermal phenomena have been seen by radio and X-ray
synchrotron emission, but more direct evidence for the existence of
the relativistic electrons in such nebulae can be given by the VHE
gamma ray radiation of inverse Compton effect.
In order to model the emission through the wide spectrum range,
more detailed studies are needed to reveal the nature of the pulsar wind
and particle acceleration, by comparing the VHE phenomena of many pulsars
 with each other.

  Radio pulsars of short periods have been also long studied
in VHE gamma ray astronomy. The binary pulsar PSR~B1259-63 with 47.8 ms
period is unique, because of its strong magnetic field, i.e., not a recycled
pulsar, and also the fact that it changes from radio pulsar to X-ray emitter
 near the periastron of binary orbital motion.
Intense stellar wind from the companion Be star presumably
 collides with the pulsar wind to generate a shock in the binary system.
 
  In this paper, we present the latest results of CANGAROO
on the two objects; PSR~B1259-63, and
 PSR~B1509-58 which is a composite system of pulsar, nebula and
supernova remnant.
 
\section{CANGAROO}
CANGAROO has confirmed the VHE gamma-ray emission from PSR1706-44 (Kifune 
et al. 1995), and also from Crab at $>$7TeV (Tanimori et al. 1994). 
Evidence of gamma-ray signals has also been detected from the direction 
of the Vela Pulsar (Yoshikoshi, 1996). 
All of three are the known gamma-ray pulsars detected by the EGRET experiment.
Another EGRET pulsar, which is in a good position for the CANGAROO 
observation, PSR1055-52 does not show the VHE gamma-ray emission above the 
CANGAROO sensitivity.
Not only the survey for the EGRET pulsars, CANGAROO is also observing many
categories of targets. 
Supernova remnant is one of the most interesting candidates.  
We have performed observations of W28 and SN1006.
And PSR1509-58 can be associated with SNR MSH15-52. 
These objects are probably extended compared with the typical spacial
resolution of the present VHE gamma-ray detectors (0.1deg).
Using the fine resolution of the CANGAROO camera, the analysis for such 
extended sources has been developed, and Vela is thought to be possibly 
extended rather than point-like. 
About SNRs, the data of SN1006 taken in 1996 indicates a VHE gamma-ray 
emission and it is probably extended. 
Observation is continuing in 1997 to confirm this indication.

The CANGAROO 3.8m telescope is located at Woomera, South Australia
(136E,31S,160m a.s.l.). 
The atmospheric \v{C}erenkov light produced by extensive air showers are 
collected by the 3.8m reflector and recorded by the imaging camera 
consisting of 256 photomultiplier tubes.
A trigger pulse is produced when any 5 PMTs exceed 3 photoelectron level.
The CANGAROO collaboration has observed targets in the southern hemisphere 
since 1992 (Hara et al., 1993).

The reflectivity of the CANGAROO 3.8m mirror was improved in October 1996 at 
the AAO (Anglo Australian Observatory), Coonabarabran.
The reflectivity of the mirror increased from 45\% to 90\% after the recoating 
and the estimated energy threshold decreased to below 1TeV at zenith.
Data shown here were taken both before and after this recoating work.

The total observation time for PSR1509-58 is shown in Table 1 and 
that for PSR1259-63 is shown in Table 2. The selected observation 
time under cloudless conditions is also shown.
Because of the different energy threshold of observations before and after 
October 1996, as mentioned above, the data are summarized in two periods for 
PSR1509-58. 
Observations of these objects are continuing in 1997, in particular PSR1259-63 
will be observed around the newest periastron (UT29.8 May97). 

%\begin{table}[h]
\begin{table}
\vspace{-6pt}
\caption{Total observation time for PSR1509-58. The selected observation time 
under cloudless conditions is also shown.}
%\label{p1509}
%%\vspace{6pt}
\begin{center}
\begin{tabular}{|l|cc|cc|}
\hline\hline
Period & On Source & Off Source & On Source selected & Off Source selected \\
\hline
May June, 1996        & 55.2 hours & 52.5 hours & 52.5 hours & 50.5 hours \\
March April May, 1997 & 42.7 hours & 41.2 hours & 36.9 hours & 35.3 hours \\
\hline
\hline
\end{tabular}
\vspace{-6pt}
\end{center}
\end{table}

%\begin{table}[h]
\begin{table}
\vspace{-6pt}
\caption{Total observation time for PSR1259-63. The selected observation time 
under cloudless conditions is also shown. Two periastrons are included during 
the observation. One is January 9, 1994 and another is May 29, 1997.}
%\label{p1259}
%%\vspace{6pt}
\begin{center}
\begin{tabular}{|l|cc|cc|}
\hline\hline
Period & On Source & Off Source & On Source selected & Off Source selected \\
\hline
January 8--10, 1994 & 8.6 hours & 4.7 hours & 8.6 hours & 4.7 hours \\
May, 1994 & 29.5 hours & 24.5 hours & 26.5 hours & 19.1 hours \\
\hline
March, 1997 & 23.1 hours & 22.4 hours & 23.1 hours & 22.4 hours \\
April, 1997 &  5.6 hours &  2.6 hours &  4.8 hours &  2.6 hours \\
\hline
\hline
\end{tabular}
\vspace{-6pt}
\end{center}
\end{table}

The analysis for the newest observations with lower energy threshold is 
in progress and results are presented at the conference.
Here we show the preliminary results of analysis on PSR1509-58 (1996 data) 
and PSR1259-63 at the previous periastron. 

\section{PSR $1509-58$}
The gamma-ray pulsar PSR1509-58 has the fifth largest \.{E}/d$^2$ among 
known radio pulsars and is the second youngest pulsar following the Crab.
Pulsed gamma-ray emission from this object was detected by the BATSE and 
OSSE experiments with a 150msec pulsar period.
Although it has not been detected by the EGRET experiment (Fierro indicates 
a DC excess with 3 sigma level), recent X-ray observations have revealed 
energetic features around this pulsar. 
From early observations, a complicated structure of X-ray nebula has been
known; the "south nebula" (SN) exists centered at the pulsar, and another
 "north nebula" (NN) at the supernova remnant MSH 15-52.
Recent ASCA observation (Tamura et al. 1996) of X-ray line emission
has clearly shown that SN is of non-thermal spectrum and NN of
thermal origin, in addition, to find emission of a jet-like shape which
connects these two nebulae. The magnetic field in the nebulae is
estimated probably to be as weak as the interstellar value.
Thus, the object is very likely a VHE gamma ray source, and it is
interesting and important to locate which part of the system may emit
inverse Compton VHE gamma rays.
du Plessis et al. (1995) performed an observation with a 7 TeV energy 
threshold and set an upper limit for VHE gamma-ray emission before the 
discovery of the jet feature.
Results of analysis for VHE gamma-ray emission from the pulsar position and
studies looking for an emission offset from the pulsar position are presented.

Each \v{C}erenkov image is fitted to an ellipse and image parameters, which 
are commonly used (e.g. Reynolds et al. 1993), are derived . 
After image selection no significant excess can be seen in the 
distribution of alpha parameter in the data of 1996.
An estimation of the 3 $\sigma$ flux upper limit from this preliminary result 
corresponds to \\

\hskip20mm F($>$3TeV)$<$ 2$\times$10$^{-12}$cm$^{-2}$s$^{-1}$ \\

\noindent
In this calculation, the source is assumed to be point-like.

An analysis assuming the thermal nebula as a source has also been performed.
No significant excess is found. This result will be compared with that 
obtained by the observation with a lower energy threshold in 1997.

\section{PSR $1259-63$}
The binary pulsar PSR1259-63 has a high eccentricity of 0.8699 
orbiting around a massive Be star (SS2883) companion with orbital period of 
1236.72 days. Be stars are known to produce substantial gaseous 
outflows (e.g. Waters et al. 1988). 
It is possible that the interaction between this gaseous nebulae and the 
pulsar relativistic wind makes a shock in this system.  
Actually some experiments have shown the emission in X-ray to soft gamma-ray
band near the previous periastron (Hirayama et al. 1996, Tavani et al. 1994, Kaspi et al. 1995).

\begin{wrapfigure}[21]{r}{8.7cm}
%\begin{center}
% \epsfile{file=fig1.eps,height=8cm,width=9cm}
   \epsfig{file=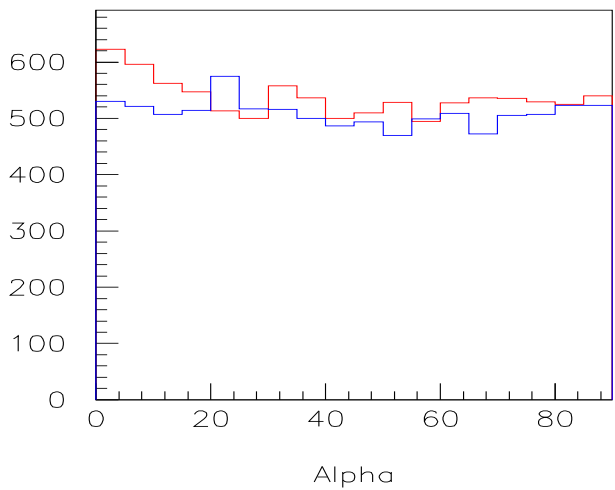,width=7.2cm,height=6cm}
\caption{ $Alpha$ $distributions$ $of$ $events$ $obtained$ 
$4$ $months$ $after$ $the$ $periastron$, $January$ $9$, $1994$, $of$ $the$ 
$PSR1259$-$63$. 
$The$ $chain$ $line$ $is$ $for$ $on$-$source$ $event$ $and$ $the$ $solid$ 
$line$ $is$ $for$ $off$-$source$ $events$. 
$Observation$ $time$ $of$ $off$-$source$ $events$ $is$ $normalized$ $to$ 
$that$ $of$ $on$-$source$ $events$. } 
\end{wrapfigure}

Gamma-ray signals were searched comparing alpha distribution between 
on-source events and off-source events. 
No significant gamma-ray signal was obtained during the previous periastron. 
On the contrary a 4.8$\sigma$ excess was obtained from the data obtained 
4 months after the periastron, although this result is preliminary.  
The gamma-ray flux is calculated preliminarily as 

F($>$3TeV) $\sim$ 4.6$\times$10$^{-12}$cm$^{-2}$s$^{-1}$

\noindent
Alpha distributions of the events in this period 
are shown in FIg. 1. These results will be compared with the 
newest results obtained in 1997 including the next periastron. 

\section{OTHER PULSARS AND DISCUSSIONS}
The established VHE gamma-ray source PSR1706-44 was also observed in 
1994 and 1995. The estimated flux from these new observations is 
consistent with the earlier result (Kifune et al. 1994). 
More observations will be performed in 1997 with a lower energy threshold. 

There still remain a lot of studies of VHE gamma rays from pulsars and 
their associated supernova remnants as is the case for the EGRET galactic 
sources, most of which are still unidentified. Observations of the 
gamma-ray sky by the CANGAROO telescope, as well as by other Air \v{C}erenkov 
Imaging Telescopes in the world, still play an important role, although 
future telescopes are planned and/or under construction.

\section{ACKNOWLEDGEMENTS}
This work is supported by International Scientific Research Program of
a Grant-in-Aid in Scientific Research of the Ministry of Education, Science,
Sports and Culture, Japan, and by the Australian Research Council.
T.~Kifune and T.~Tanimori acknowledge the support of the Sumitomo Foundation.
The receipt of JSPS Research Fellowships (P.G.E., T.N., M.D.R., K.S., G.J.T.
and T.Yoshikoshi) is also acknowledged.

\section{REFERENCES}
\setlength{\parindent}{-5mm}
\begin{list}{}{\topsep 0pt \partopsep 0pt \itemsep 0pt \leftmargin 5mm
\parsep 0pt \itemindent -5mm}
\vspace{-15pt}

\item du Plessis, I., et al., ApJ, 453, 746 (1995).
% \item Punch, M., Ph.D. thesis, National University of Ireland (1993).
\item Fierro, J.M., Ph.D. thesis, Stanford University (1995).
\item Hara, T., et al., Nucl. Instr. Meth., A332, 300 (1993).
\item Hirayama, M., et al., Pub. Astron. Soc. Japan 48, 833 (1996).
\item Kaspi, V.M., et al., ApJ, 453, 424 (1995).
\item Kifune, T., et al. ApJ, 438, L91 (1995).
\item Reynolds, P.T., et al., ApJ, 404, 206 (1993).
\item Tamura, K., et al., PASJ, 48, L33 (1996). 
\item Tanimori, T., et al., ApJ, 429, L61 (1994).
\item Tavani, M., et al., BAAS, 185, 10,214 (1994).
\item Waters, L.B.F.M., et al., A\&AS, 198, 200 (1988).
\item Yoshikoshi, T., Ph.D. thesis, Tokyo Institute of Technology (1996).

% \item Bailes, M., {\it et al.} 1989, \apjl, 343, L53
% \item Bhat, P.N., et al., A\&A, 178, 242 (1987).
% \item Buccheri, R., et al., A\&A, 128, 245 (1983).
% \item de~Jager, O.C., Harding, A.K. \& Strickman, M.S. 1996,
% \item Edwards, P.G., et al., A\&A, 291, 468 (1994).
% \item Grindlay, J.E., et al. ApJ, 201, 82 (1975).
%                     \apj, 460, 729
% \item Harnden, F.R., {\it et al.} 1985, \apj, 299, 828
% \item Hillas, A.M. 1985, Proc. 19th Int. Cosmic Ray Conf. (La Jolla), 
%                     3, 445
% \item Kanbach, G., {\it et al.} 1994, \aap, 289, 855
% \item Kennel, C.F. \& Coroniti, F.V. 1984, \apj, 283, 694
% \item Markwardt, C.B. \& \"{O}gelman, H. 1995, \nat, 375, 40
% \item Markwardt, C.B. \& \"{O}gelman, H. 1997, \apjl, 480, L13
% \item \"{O}gelman, H., Koch-Miramond, L. \& Auri\'{e}re, M.
%                     1989, \apjl, 342, L83
% \item \"{O}gelman, H., Finley, J.P. \& Zimmermann, H.U. 1993,
%                     \nat, 361, 136
% \item Punch, M., Ph.D. thesis, National University of Ireland (1993).
% \item Weekes, T.C., et al., ApJ, 342, 379 (1989).
% \item Yoshikoshi, T. et al., ApJ, in preparation (1997).

\end{list}

\end{document}